\documentclass[12pt]{article}
\usepackage{float}
\usepackage{amsmath,amssymb}
\usepackage{mathrsfs}
\usepackage{braket}
\usepackage[dvipdfmx]{color}
\usepackage{framed}
\usepackage{graphicx}

\usepackage{bm}
\usepackage{subcaption}
\usepackage{cite}

\makeatletter

\renewcommand\section{\@startsection {section}{1}{\z@}%
                                   {-3.5ex \@plus -1ex \@minus -.2ex}%
                                   {2.3ex \@plus.2ex}%
                                   {\normalfont\large\bfseries}}

\renewcommand\subsection{\@startsection{subsection}{2}{\z@}%
                                     {-3.25ex\@plus -1ex \@minus -.2ex}%
                                     {1.5ex \@plus .2ex}%
                                     {\normalfont\normalsize\bfseries}}

\makeatother

\usepackage[driverfallback=dvipdfmx]{hyperref}
\hypersetup{%
  setpagesize=false,
  bookmarksnumbered=true,
  bookmarksopen=true,
  colorlinks=true,
  linkcolor=blue,
  citecolor=blue,
  linktoc=all, 
  linktocpage=true, 
  }

\begin{document}

\baselineskip=18pt  
\numberwithin{equation}{section}  
\allowdisplaybreaks  



%
%


\thispagestyle{empty}

\vspace*{-2cm}

\begin{center}

\vspace{1.4cm}

{\bf \Large On Stabilization of Magnetically Charged Brane Shell\\ and Over-extremality} 
\vspace*{0.2cm}

\vspace{1.3cm}


{\bf
Sohei Tsukahara$^{1}$} \\
\vspace*{0.5cm}

${ }^{1}${\it Department of Physics, Kyushu University, Fukuoka 810-8581, Japan  }\\

\vspace*{0.5cm}

\vspace*{0.5cm}

\end{center}

\vspace{1cm} \centerline{\bf Abstract} \vspace*{0.5cm}

In string theory, we can geometrically realize a metastable state by wrapping D5-branes and anti D5-branes to a singular manifold. We consider wrapping D3-branes to the internal space in this setup. These D3-branes dissolve into the domain wall, which interpolates true vacua and false vacua, forming a bound state. The remnant of the D3-branes can be seen as a background magnetic field on the domain wall, which appears to an observer in 4D spacetime as a magnetically charged, spherically symmetric shell. This brane shell has finite radii due to the nonlinearity peculiar to string theory, even at the probe level. We demonstrate a new stabilization mechanism of the brane shell in 4D spacetime. We add a general relativity-inspired gravitational correction to the brane shells and investigate the influence on its potential. As a result, the potential value at the horizon will be relatively larger than the potential minimum in parameter regions where the influence of gravity is large, and even non-perturbative instabilities can be removed. Moreover, we show the existence of over-extremal states such that $(gQ)^{2}\geq G_{4}m^{2}$ is satisfied in regions where the magnetic field is sufficiently large. At least in our model, this over-extremal shell cannot be completely stabilized by gravitational correction. This paper also addresses the dilemma between stabilizing and achieving an over-extremal state.

\newpage
\setcounter{page}{1} 

\newpage
\setcounter{page}{1} 



\tableofcontents
\vspace{0.5cm}

\section{Introduction}

Although string theory is considered one of the best candidates for quantum gravity, its vacuum structure is considered to be complicated, and there have yet to be successful attempts to derive a unique, effective theory in a top-down way. Given this situation, a bottom-up approach that expresses a mathematical condition to be satisfied by a UV-complete low-energy effective theory has recently attracted attention. This attempt is called the Swampland problem, and various conjectures have been proposed based on the empirical rules of unified theory and early universe \cite{Vafa:2005ui,Ooguri:2006in,Ooguri:2016pdq,Obied:2018sgi,Garg:2018reu,Ooguri:2018wrx,Lust:2019zwm}. There are many reviews on this field such as \cite{Palti:2019pca,Brennan:2017rbf,Danielsson:2018ztv,Akrami:2018ylq,Grana:2021zvf,vanBeest:2021lhn}.

Weak Gravity Conjecture (WGC)\cite{Arkani-Hamed:2006emk} is one of the famous swampland conjectures, which gives a lower bound to the gauge coupling constant (see, e.g., \cite{Palti:2020mwc,Harlow:2022ich} for reviews). In the original version, this conjecture states that there must be at least one particle that satisfies an inequality 
\begin{align}
    \left(gQ \right)^{2}\geq \frac{m^{2}}{2M_{\rm pl}^{2}}=G_{4}m^{2}\ . \label{WGcond-original}
\end{align}
for $U(1)$ gauge coupling\footnote{This statement is equivalent to the claim that black holes not protected by (super)symmetry must necessarily decay. If we apply this to extremal black holes, there may exist WGC states where the mass-to-charge ratio is less than one due to quantum corrections. This tendency has been confirmed by various gravitational theories. For discussions in the context of Einstein-Maxwell theory, see\cite{Natsuume:1994hd,Kats:2006xp,Cheung:2018cwt,Hamada:2018dde,Bellazzini:2019xts,Charles:2019qqt,Jones:2019nev,Loges:2019jzs,Goon:2019faz,Cano:2019oma,Cano:2019ycn,Cremonini:2019wdk,Chen:2020rov,Loges:2020trf,Bobev:2021oku,Arkani-Hamed:2021ajd,Ma:2021opb,Cremonini:2021upd,Aalsma:2021qga,Noumi:2022ybv}, and for recent discussions extending to gravitational theories coupled with nonlinear electrodynamics, see\cite{Abe:2023anf}.}. Such particle is often said to be {\it over-extremal}. In theories decoupled from gravity, the right-hand side is zero, a trivial inequality. However, the coupling constant cannot be completely zero once we consider an effective theory coupled with gravity. Whether or not such a compact object can be constructed in the framework of quantum gravity (e.g., string theory) is a nontrivial, important problem.

One of the candidates considered in string theory is a bubble-like object formed by Dp-branes. This paper refers to such objects as {\it brane shells}. The formation of brane shells is a well-known phenomenon, exemplified by the Myers effect, where D0-branes condense and expand into a fuzzy two-sphere \cite{Myers:1999ps}.\footnote{Another pioneering study on Dp-brane polarization is \cite{Emparan:1997rt}.} Generally speaking, some mechanisms to stabilize the system is necessary to keep the brane shells at a finite radius. In the case of the Myers effect, the polarized brane can be considered supported by the background Ramond-Ramond flux. One can see other instances of bubble generation through brane/flux annihilation in \cite{Kachru:2002gs,Gautason:2015tla}.

Additionally, in recent years, a new example of polarized brane shells has been presented in discussions about black holes \cite{Danielsson:2022odq}. The quantum picture of black holes is still unclear, and various top-down hypotheses, such as the fuzzball proposal \cite{Mathur:2002ie,Mathur:2005zp,Bena:2022rna}, have been proposed. One of these is the so-called {\it AdS bubble proposal}, which states that a black hole is a spherical shell with an internal supersymmetric AdS space and an external metastable vacuum\cite{Danielsson:2017riq,Danielsson:2017pvl} (see also \cite{Danielsson:2021ykm,Danielsson:2021ruf,Danielsson:2023onu,Giri:2024cks} for related studies). The authors of \cite{Danielsson:2022odq} discussed the stringy embedding of this proposal and constructed a counterpart to the Reissner-Nordstr\"{o}m black hole under Type IIB compactification. Their brane shell is charged due to the Freed-Witten effect \cite{Freed:1999vc}, and in a situation where there are sufficiently many D3-branes wrapped in the internal space, it is indeed over-extremal.\footnote{Such solutions representing over-extremal shells have also been proposed in general relativity in recent years \cite{Kehagias:2023qmy}.}

In this paper, we realized a metastable state geometrically with D5-branes and anti D5-branes and considered a situation that a brane shell via a domain wall D5-brane is nucleated in 4D spacetime\cite{Cachazo:2001jy,Vafa:2000wi,Aganagic:2006ex}. If we wrap D3-branes into the internal space, D3-branes dissolve into (anti) D5-branes and construct a bound state\cite{StringText1,StringText2}. When viewed in 4D spacetime, this bound state can be observed as a magnetically charged brane shell. In this situation, we add a GR-inspired gravitational correction to the shell and examine the stability against tunneling and the over-extremality. As a result, we found a mechanism for stabilizing the brane shell due to the gravitational effect and confirmed numerically that the over-extremal state is realized for enough magnetic field strength. In addition, compared to the case in \cite{Danielsson:2022odq}, it was confirmed that if the magnetic field was taken too large, the instability would be so strong that the shell could not exist, or the theory would be so meaningless that physical spacetime would not exist. We also discuss this upper bound on the magnetic field. This study was inspired by \cite{Danielsson:2022odq}.

The remainder of this paper is organized as follows: in section  \ref{Review of bound state between domain wall D5 and monopole}, we review the bound state between the domain wall D5-brane and D3-branes. In section \ref{Stabilization of brane shell}, we discuss the stability of the brane shell under gravitational correction and its over-extremal property. Section \ref{Discussion and summary} is devoted to a summary and discussion.

\section{Review of bound state between domain wall D5 and monopole}
\label{Review of bound state between domain wall D5 and monopole}

Let us consider a noncompact Calabi-Yau threefolds which satisfies \cite{Cachazo:2001jy,Vafa:2000wi,Aganagic:2006ex}
\begin{align}
\label{InternalSpace}
0=z_1^2+z_2^2+z_3^2 +W^{\prime}(z_4)^2, \qquad W^{\prime}(z_4) = g (z_4-a_1)(z_4-a_2),
\end{align}
where $z_{i}$ are complex variables and nondimensionalized by string length $l_{s}$. This threefold is apparently singular at $z_{4}=a_{1}$ and $z_{4}=a_{2}$, but we can resolve it by replacing it with a finite two-sphere and blowing up. Here, we wrap $N_{1}$ D5-branes to a two-cycle $[C_{1}]$ at $z_{4}=a_{1}$ and $N_{2}$ anti D5-branes to another two-cycle $[C_{2}]$ at $z_{4} = a_{2}$. As $[C_{1}]$ and $[C_{2}]$ belong to a same homology class, D5-branes and anti D5-branes annihilate with each other. However, they must gain additional energy to climb over the $\mathbb{S}^{3}$ between them. This means that there is some potential barrier between the two branes. In this way, the supersymmetry-broken metastable state is geometrically realized. In the following, we assume $N_{1}=N_{2}=1$ for simplicity.

Next, we wrap D3-branes to $\mathbb{S}^{3}$ in the internal space\cite{Kasai:2015exa,Tsukahara:2023xzy}. These wrapped D3-branes dissolve into a domain wall D5-brane, which intermediates between a true vacuum and false vacuum and forms a bound state. This is because it is an energetically more favorable state rather than the stand-alone \cite{StringText1,StringText2}:
\begin{align}
    E_{D3}+E_{D5} > E_{\rm bound} = \sqrt{E_{D3}^{2}+E_{D5}^{2}}.
\end{align}
We can see the remnant of dissolved D3-branes as background magnetic flux on the domain wall D5-brane. That is, from the observer's viewpoint in 4D external space, the bound state of a monopole and the domain wall bubble appear to be realized.

In the above setup, the magnetic field plays a role as a catalyst to enhance the instability of the brane shell. We shall derive the potential of the brane shell to confirm the instability. Since the main contribution to the potential is given by Dirac-Born-Infeld (DBI) energy, we have to calculate the DBI action of the domain wall D5-brane and the (anti) D5-brane, respectively. The induced metric on the D5-brane is given by \cite{Cachazo:2001jy,Vafa:2000wi,Aganagic:2006ex}
\begin{align}
    \label{inducedMetric-D5}
    \frac{ds^{2}_{\rm D5/\overline{D5}}}{l_{s}^{2}}=-dT^{2}+d\xi^{2}+\xi^{2}\left(d\theta^{2}+\sin^{2}\theta d\varphi^{2} \right)+L^{2}\sin^{2}\psi_{I}\left(d\theta_{I}^{2}+\sin^{2}\theta_{I}d\varphi^{2}_{I} \right)  , 
\end{align}
where $T$, $\xi$ and $L$ are dimensionless variables. The background magnetic flux as the remnant of the D3-brane exists in the $(\theta,\varphi)$ direction, so the  $U(1)$ field strength part in the DBI action should be $2\pi \alpha^{\prime}F_{\theta \varphi}=b_{D3}\sin\theta$. This $b_{D3}$ is proportional to the number of D3-branes: $b_{D3}\propto \#_{D3}$. In addition, we also have a NS B field $B_{\theta_{I}\varphi_{I}}^{NS}$ as a background field in $(\theta_{I},\varphi_{I})$ direction of the internal space, where the subscription ${}_{I}$ denotes the internal coordinates. At this point, we can calculate the determinant factor of the DBI action as 
\begin{align}
    \label{detFctor-D5DBI}
    -\mathrm{det}\left(G_{ab}+B_{ab}+2\pi \alpha^{\prime}F_{ab} \right) = \sin^{2}\theta\left(\xi^{4}+b_{D3}^{2} \right) \times\left(L^{4}\sin^{4}\psi_{I}\sin^{2}\theta_{I} + (B_{\theta_{I} \varphi_{I}}^{NS})^{2} \right) .
\end{align}
The singular points where the D5-brane and anti D5-brane are located, $z_{4} = a_{1}$ and $z_{4} = a_{2}$, correspond to the internal space coordinates $\psi_{I} = 0$ and $\pi$. Then, taking $\sin\psi_{I} = 0$, the DBI Lagrangian of the (anti) D5-brane yields
\begin{align}
    \label{lagrangian-D5}
    L_{D5}=T_{D5} r \left(\int^{\infty}_{0}-\int^{\infty}_{R} \right) d\xi \Big( 4\pi \sqrt{\xi^4+b_{D3}^2}\Big)=T_{D5} r  b_{D3} \Big[4\pi R\cdot {}_2F_{1} \Big(-{1\over 2}, {1\over 4}, {5\over 4},-{R^4\over b_{D3}^2} \Big) \Big] ,
\end{align}
where $R$ is the radius of the $\mathbb{S}^{2}$ bubble in the 4D spacetime, and $r$ is the integrated value of $B$ field: $r=\int_{\mathbb{S}^{2}}B_{2}^{NS}$. ${}_{2}F_{1}$ is a generalized hypergeometric function. In this calculation, we adopt as our energy reference the configuration in which the 4D spacetime is filled with (anti) D5-branes, i.e., the situation in which supersymmetry is restored.

On the other hand, the induced metric on the domain wall D5-brane that constructs a brane shell in the 4D spacetime is given by
\begin{align}
    \label{inducedMetric-dD5}
    \frac{ds^{2}_{\rm DWD5}}{l_{s}^{2}}=-dT^{2}+R^{2}\left(d\theta^{2}+\sin^{2}\theta d\varphi^{2} \right)+L^{2}\left[d\psi^{2}_{I}+\sin^{2}\psi_{I}\left(d\theta^{2}_{I}+\sin^{2}\theta_{I}d\varphi^{2}_{I} \right) \right] .
\end{align}
Since the configurations of the magnetic flux and the B field are the same as the case of (anti) D5-brane, calculating the determinant factor, the DBI Lagrangian of the domain wall D5-brane is expressed as follows
\begin{align}
    \label{lagrangian-dD5}
    L_{ DW}= -T_{DW} 4\pi   \sqrt{(R^4+b_{D3}^2)(1-\dot{R}^2)}. 
\end{align}
In the calculation, we define the tension of the domain wall as
\begin{align}
    T_{DW}=T_{D5}\Big[2\pi^2 L^3 \int_0^\pi d \psi_I \, {2\over \pi}\sqrt{\sin^4 \psi_I +\Big({b_{NS}\over L^2}\Big)^2 } \Big] ,
\end{align}
where $b_{NS}=r/4\pi$. Combining \eqref{lagrangian-D5} and \eqref{lagrangian-dD5}, we obtain the Lagrangian for the brane shell \cite{Tsukahara:2023xzy}
\begin{align}
L_{\rm total}&= L_{DW}  +2\times L_{D5} \notag \\
&= -T_{DW} 4\pi   \sqrt{(R^4+b_{D3}^2)(1-\dot{R}^2)}+2T_{D5} r  b_{D3} \Big[4\pi R\cdot {}_2F_{1} \Big(-{1\over 2}, {1\over 4}, {5\over 4},-{R^4\over b_{D3}^2} \Big) \Big] . \label{probeLag}
\end{align}

\begin{figure}[tb]
    \centering
    \includegraphics[width=0.5\textwidth]{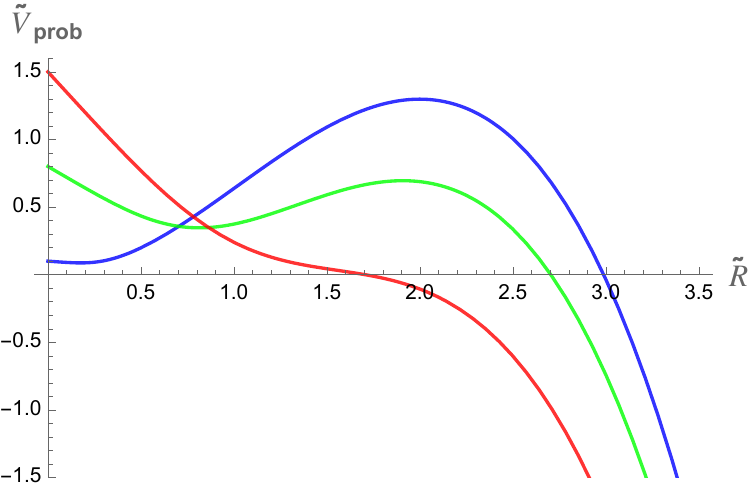}
    \caption{Probe level potentials for each strength of the magnetic field. Blue line for $\widetilde{b}_{D3}=0.1$, green line for $\widetilde{b}_{D3}=0.8$ and red line for $\widetilde{b}_{D3}=1.5$. The larger the magnetic field, the lower the potential barrier, and the more decay is promoted. The range of dimensionless magnetic field in which the extreme values of the potential exist is $0\leq \tilde{b}_{D3} <\frac{3\sqrt{3}}{4} \simeq 1.299\cdots$.
    }
    \label{StaticEnergy_D5antiD5}
\end{figure}

Imposing $\dot{R}=0$ to the Lagrangian \eqref{probeLag} and reversing the overall sign, we get the potential of the brane shell at the probe level as 
\begin{align}
    V_{\rm prob} = T_{DW} 4\pi   \sqrt{R^4+b_{D3}^2} - 2T_{D5} r  b_{D3} \Big[4\pi R\cdot {}_2F_{1} \Big(-{1\over 2}, {1\over 4}, {5\over 4},-{R^4\over b_{D3}^2} \Big) \Big] . \label{probPot}
\end{align}
This potential has a potential barrier whose height depends on the strength of the magnetic field, suggesting that tunneling can occur in a finite lifetime (see Figure \ref{StaticEnergy_D5antiD5}).\footnote{See \cite{Tsukahara:2023xzy} for a complete one-loop analysis of this model.} It should be noted that due to the nonlinearity of the DBI action, the potential has a minimum value at nonzero $R$. In other words, even before tunneling occurs, there appears to be a brane shell of finite size to an observer in 4D spacetime.

\section{Stabilization of brane shell}
\label{Stabilization of brane shell}

\subsection{Gravitational correction via junction condition}
\label{Gravitational correction via junction condition}

Following the discussion in \cite{Danielsson:2017riq,Danielsson:2022odq}, we add a gravitational correction to the brane shell potential \eqref{probPot}. The authors of \cite{Danielsson:2017riq,Danielsson:2022odq} have considered that the interior of the shell is a stable AdS spacetime and the exterior is a metastable vacuum. If we denote the outside of the shell by $+$ and the inside by $-$, the metric of each spherically symmetric spacetime can be expressed as 
\begin{align}
    ds^{2}_{\pm}=-f_{\pm}(r)dt^{2}+\frac{dr^{2}}{f_{\pm}(r)}+r^{2}d\Omega_{2}^{2} ,\label{2.15}
\end{align}
where $d\Omega_{2}^{2}$ is the standard 2D round metric. In this situation, as the lapse function only depends on $r$, the Israel junction condition takes \cite{Israel:1966rt}
\begin{align}
    \mathcal{T}(r)=\frac{2}{8\pi G_{4}}\frac{1}{r}\left(\sqrt{f_{-}(r)}-\sqrt{f_{+}(r)} \right) ,\label{Dani2.18}
\end{align}
where $\mathcal{T}(r)$ is the tension of the shell.  In the later part, we clarify the relation between this tension and the aforementioned domain wall tension.

First, we need to choose the interior and exterior spacetime properly to add a gravitational correction. In the brane setup shown in the section \ref{Review of bound state between domain wall D5 and monopole}, we considered the configuration in which the 4D spacetime is filled with (anti) D5-branes as energy reference. Thus, when adding the correction, it would be plausible to see the inner spacetime (true vacuum) as the Minkowski vacuum, which has zero gravitational energy. That is, $f_{-}(r) = 1$ holds. In contrast, we must consider the Einstein gravity coupled to a nonlinear electromagnetic Lagrangian outside the shell (false vacuum). This point requires some cautions. The well-known nonlinear electromagnetic Lagrangian called DBI Lagrangian is given by the following expression \cite{Born:1934gh,Dirac:1962iy}
\begin{align}
    \mathcal{L}_{DBI}&=-\frac{1}{\gamma}\left[\sqrt{-\operatorname{det}\left(\eta_{\mu \nu}+\sqrt{\gamma}F_{\mu \nu}\right)}-\sqrt{-\operatorname{det}\left(\eta_{\mu \nu}\right)} \right] \nonumber \\
    &=-\frac{1}{\gamma}\left[\sqrt{1+\frac{\gamma}{2}F_{\mu \nu}F^{\mu \nu}-\frac{\gamma^{2}}{16}\left(F_{\mu \nu}\widetilde{F}^{\mu \nu} \right)^{2}}-1 \right]\ ,\label{BIlag}
\end{align}
where $\gamma$ is the Born-Infeld parameter, determining the cutoff scale, and $\widetilde{F}^{\mu \nu}$ is the dual tensor defined as $\widetilde{F}^{\mu \nu}=(1/2)\epsilon^{\mu \nu \alpha \beta}F_{\alpha \beta}$. If $F_{\mu \nu}$ is tiny, this Lagrangian becomes the usual Maxwell Lagrangian. However, in the present case, it is preferable to attribute it to the Nambu-Goto Lagrangian when $F_{\mu \nu}\rightarrow0$. Thus, we should adopt
\begin{align}
    \mathcal{L}^{\prime}_{DBI}=-\frac{1}{\gamma}\sqrt{1+\frac{\gamma}{2}F_{\mu \nu}F^{\mu \nu}-\frac{\gamma^{2}}{16}\left(F_{\mu \nu}\widetilde{F}^{\mu \nu} \right)^{2}}\ \label{BIlike}
\end{align}
as the appropriate Lagrangian, which couples to the Einstein gravity outside the shell.

To derive a brane shell potential, all we have to do is to add energies of the inner region of the shell, the outer region of the shell, and the shell itself. As we mentioned above, the first one is zero. For the second one, the energy-momentum tensor of the gravity coupled to \eqref{BIlike} is given by \cite{Bronnikov:2000vy,Kruglov:2017mpj}
\begin{align}
    T_{\mu \nu}=-\Pi^{-1/2}\left(F_\mu{ }^\alpha F_{\nu \alpha}-\gamma \mathcal{G} \widetilde{F}_\mu{ }^\alpha F_{\nu \alpha}\right)-g_{\mu \nu} \mathcal{L}\ ,\quad \Pi=1+\frac{\gamma}{2}F_{\alpha \beta}F^{\alpha \beta}-\frac{\gamma^{2}}{16}\left(F_{\alpha \beta}\widetilde{F}^{\alpha \beta} \right)^{2} ,
\end{align}
where $\mathcal{G}=(1/2)F_{\mu \nu}\widetilde{F}^{\mu \nu}$. Substituting the Lagrangian \eqref{BIlike} into this, we obtain
\begin{align}
    T_{\mu \nu}&=-\left(1+\frac{\gamma}{2}F_{\alpha \beta}F^{\alpha \beta} \right)^{-1/2}{F_{\mu}}^{\alpha}F_{\nu \alpha}+\frac{g_{\mu \nu}}{\gamma}\sqrt{1+\frac{\gamma}{2}F_{\alpha \beta}F^{\alpha \beta}} \nonumber \\
    &=-\left(1+\frac{\gamma}{2}F_{\alpha \beta}F^{\alpha \beta} \right)^{-1/2}\left\{{F_{\mu}}^{\alpha}F_{\nu \alpha}-\frac{g_{\mu \nu}}{\gamma}\left(1+\frac{\gamma}{2}F_{\alpha \beta}F^{\alpha \beta} \right) \right\} . \label{energyMomentum}
\end{align}
We already omitted $F\widetilde{F}$ term in the first line because our setup has no electric field. Since we are assuming a spherically symmetric spacetime, let us impose the following ansatz on the metric
\begin{align}
    ds^{2}=-f(r)dt^{2}+f^{-1}(r)dr^{2}+r^{2}\left(d\theta^{2}+\sin^{2}\theta d\phi^{2} \right) \ .
\end{align}
Putting this into \eqref{energyMomentum}, we have the energy density of the spacetime as
\begin{align}
    \rho_{M}&={T^{0}}_{0} \nonumber \\
    	&=f^{-1}(r)\left(1+\frac{\gamma}{2}F_{\alpha \beta}F^{\alpha \beta} \right)^{-1/2}\left\{{F_{0}}^{\alpha}F_{0\alpha}+\frac{f(r)}{\gamma}\left(1+\frac{\gamma}{2}F_{\alpha \beta}F^{\alpha \beta} \right) \right\}. \nonumber  \\
	&=\gamma^{-1}\sqrt{1+\frac{\gamma}{2}F_{\alpha \beta}F^{\alpha \beta}} \nonumber \\
    	&=\gamma^{-1}\sqrt{1+\gamma \frac{q^{2}}{r^{4}}} \ .\label{T00}
\end{align}
We have dropped ${F_{0}}^{\alpha}F_{0\alpha}$ in the third line because there is no electric field. Also, since the magnetic field does exist in $r$-direction, we used 
\begin{align}
    F_{\mu \nu}F^{\mu \nu}=F_{23}F^{23}+F_{32}F^{32}=\frac{2q^{2}}{r^{4}}\ ,
\end{align}
where $F_{23}=q\sin\theta$. Naively, it seems that we should integrate the energy density $\rho_{M}$ by $(r,\infty)$, but as we stated so far, we took the vacuum with supersymmetry restored as the energy reference on the string theory side, so $4\pi \int^{\infty}_{0}\rho_{M}(r)r^{2}dr$ should be subtracted as a background. That is, 
\begin{align}
    V_{+}&=4\pi \left(\int^{\infty}_{r}-\int^{\infty}_{0} \right)drr^{2}\rho_{M}(r) \nonumber \\
    &=-4\pi \int^{r}_{0}drr^{2}\rho_{M}(r) \nonumber \\
    &=-4\pi\gamma^{-1/2} rq {}_{2}F_{1}\left(-\frac{1}{2},\frac{1}{4},\frac{5}{4};-\frac{r^{4}}{\gamma q^{2}} \right)\ .\label{BIenergy-0r}
\end{align}

Next, let us consider the shell's energy. In our model, since the domain wall D5-brane forms the shell, its energy can be derived by multiplying the tension by the surface area. This tension is related to the action of the domain wall D5-brane as follows
\begin{align}
    \frac{S_{\rm DW}}{2}=\mathcal{T}(r)\int \sqrt{-h}d^{3}x,
\end{align}
where $h$ is a determinant of an induced metric on the shell. Here, the $1/2$ factor on the left side comes from the fact that the current discussion focuses on a single shell, whereas the probe-level Lagrangian describes two bubbles arising at $z_{4}=a_{1}$ and $a_{2}$. From this equation, we can find the following relationship between the tension and the domain wall potential
\begin{align}
 4\pi \mathcal{T}(r)r^{2}dt=\frac{V_{\rm DW}}{2}dT=2\pi T_{\rm DW}\sqrt{R^{4}+b^{2}_{\rm D3}}dT.
\end{align}
This equation implies that we can write the shell's tension as 
\begin{align}
    \mathcal{T}(r)=\frac{T_{\rm DW}}{2r^{2}}\sqrt{r^{4}+b^{2}_{\rm D3}} . \label{shellTension}
\end{align}
The junction condition shown in \eqref{Dani2.18} then becomes 
\begin{align}
    \frac{T_{\rm DW}}{2r^{2}}\sqrt{r^{4}+b^{2}_{\rm D3}}=\frac{2}{8\pi G_{4}}\frac{1}{r}\left(\sqrt{f_{-}(r)}-\sqrt{f_{+}(r)} \right) \ .
\end{align}
Multiplying $\mathcal{T}$ in \eqref{shellTension} by the surface area of the shell and $\frac{1}{2}(\sqrt{f_{-}}+\sqrt{f_{+}})$, we obtain the contribution of the shell as
\begin{align}
    V_{\rm shell} = 4\pi r^{2}\mathcal{T}(r) \cdot\frac{1}{2}\left(\sqrt{f_{-}(r)}+\sqrt{f_{+}(r)} \right). \label{shellEne}
\end{align}

Combining the results of \eqref{BIenergy-0r} and \eqref{shellEne}, we obtain the brane shell potential as
\begin{align}
    V_{\rm tot} 
    &= V_{-} + V_{\rm shell} + V_{+} \nonumber \\
    &= 4\pi r^{2}\mathcal{T}(r) \cdot\frac{1}{2}\left(\sqrt{f_{-}(r)}+\sqrt{f_{+}(r)} \right) -4\pi\gamma^{-1/2} rq {}_{2}F_{1}\left(-\frac{1}{2},\frac{1}{4},\frac{5}{4};-\frac{r^{4}}{\gamma q^{2}} \right) .\label{totalEnergy}
\end{align}
By comparing this expression with the probe-level potential \eqref{probPot}, we can read the following relations for the parameters
\begin{align}
    q=\gamma^{-1/2}b_{D3}\ ,\quad \gamma^{-1}=T_{ D5}r_{NS}\ ,\label{probgrav-coresp}
\end{align}
where we denote the integrated value of the B field on the cycle as $r_{NS}$ to distinguish from the coordinate. Finally, the total potential with the gravitational correction can be written using $b_{D3}$ and $T_{D5}$ as follows
\begin{align}
	V_{\rm tot} = 2\pi T_{\rm DW} \sqrt{r^{4}+b^{2}_{\rm D3}}\cdot\frac{1}{2}\left(\sqrt{f_{-}(r)}+\sqrt{f_{+}(r)} \right) -4\pi T_{D5}r_{NS}b_{D3}r {}_{2}F_{1}\left(-\frac{1}{2},\frac{1}{4},\frac{5}{4};-\frac{r^{4}}{b_{D3}^{2}} \right) , \label{totalEnergy_rev}
\end{align}
In the following part, we will use \eqref{totalEnergy_rev} to discuss the stability of the brane shell.

\begin{figure}[tb]
    \centering
    \includegraphics[width=0.5\linewidth]{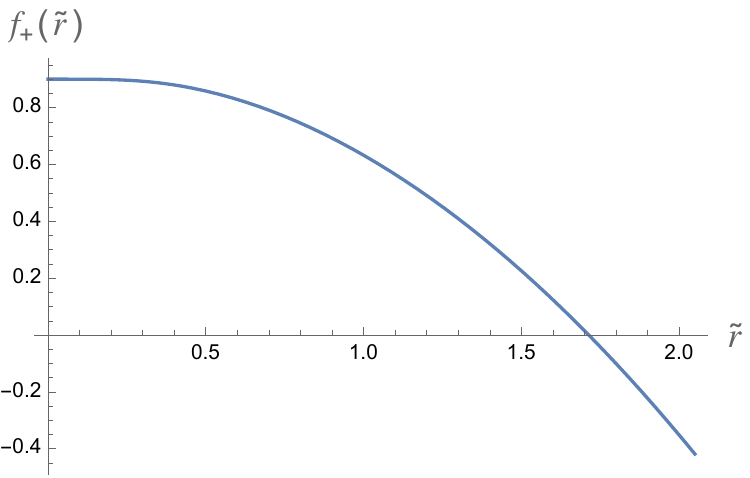}
    \caption{A lapse function at $\widetilde{b}_{D3}=0.1$ and $\widetilde{G}_{4}=0.5$. There is indeed a ``cosmological horizon'' at $r\approx 1.7$.
    }
    \label{LapseEg}
\end{figure}

At the end of this subsection, let us determine $f_{+}(r)$. The Einstein equation in a spherical spacetime is given by \cite{Wald:1984rg,Poisson:2009pwt}
\begin{equation}
    \begin{gathered}
        \frac{\partial m}{\partial r}=4 \pi r^2\left(-{T^{0}}_{0}\right), \quad \frac{\partial m}{\partial t}=-4 \pi r^2\left(-{T^{1}}_{0}\right), \\
        \frac{\partial \psi}{\partial r}=4 \pi r f^{-1}\left(-{T^{0}}_{0}+{T^{1}}_{0}\right) .
    \end{gathered}
    \label{5.4}
\end{equation}
Since we think about static spacetime, solving the first equation to obtain the mass function is enough. Substituting \eqref{T00} into ${T^{0}}_{0}$, we have
\begin{align}
    \frac{\partial m}{\partial r}&=-\frac{4\pi r^{2}}{\gamma}\sqrt{1+\gamma \frac{q^{2}}{r^{4}}} = -4\pi r^{2} T_{D5}r_{NS}\sqrt{1+\frac{b^{2}_{D3}}{r^{4}}} \nonumber \\
    \therefore m(r)&=-4\pi T_{D5}r_{NS} \left(\int^{\infty}_{r} - \int^{\infty}_{0} \right)dr\sqrt{r^{4}+b_{D3}^{2}} 
    = 4\pi T_{D5}r_{NS}rb_{D3} {}_{2}F_{1}\left(-\frac{1}{2},\frac{1}{4},\frac{5}{4};-\frac{r^{4}}{b_{D3}^{2}} \right)\ .
\end{align}
Thus, the expression of $f_{+}(r)$ is 
\begin{align}
    f_{+}(r)=1-\frac{2G_{4}m(r)}{r}\ ,\quad m(r)=4\pi T_{D5}r_{NS}b_{D3} {}_{2}F_{1}\left(-\frac{1}{2},\frac{1}{4},\frac{5}{4};-\frac{r^{4}}{b_{D3}^{2}} \right) .
\end{align}
This $f_{+}(r)$ is the monotonically decreasing function with respect to $r$, so it takes the maximum value at $r=0$ and zero at $r=r_{\rm ch}$, see Figure  \ref{LapseEg}. There is no longer a physical spacetime in the region of $r$ larger than $r_{\rm ch}$; there is a ``cosmological horizon'' in the spacetime outside the brane shell. This is an important fact when evaluating the stability of the shell in the next section. A necessary condition for the existence of this horizon is that the maximum value of $f_{+}$ is greater than zero. Namely, 
\begin{align}
    f_{+}(0) = 1-8\pi G_{4}b_{D3}T_{D5}r_{NS} > 0 . \label{horizonCond}
\end{align}
The position of the horizon is determined by the brane tension, the Newton constant, and the magnetic field, and we can obtain the numerical value by solving $f_{+}(r_{\rm ch}) = 0$.

\subsection{Stabilization mechanism}
\label{Stabilization mechanism}

The fact that the shell has a finite radius is equivalent to the potential having a nonzero minimum value. To have such a minimum, $r^{\star}$, which satisfies $\partial V_{\rm tot}/\partial r^{\star}=0$, needs to exist. Strictly speaking, we have to show that the second derivative of the potential at $r=r^{\star}$ is positive, but since the derivative at $r=0$ is always negative in our case
\begin{align}
    \frac{\partial V_{\rm tot}(0)}{\partial r} = -\left.4\pi T_{D5}r_{NS}\sqrt{r^{4} +b_{D3}^{2}} \right|_{r=0}
    = -4\pi b_{D3}T_{D3}r_{NS} < 0,
\end{align}
we have at least one minimum if $r^{\star}$ exists. At this time, it is impossible to explicitly write down $r^{\star}$ because $f_{+}$ is written by the hypergeometric function. So, in practice, we have no choice but to rely on numerical calculations.

\begin{figure}[tb]
\begin{minipage}[t]{0.5\hsize}
\centering
\includegraphics[width=0.9\linewidth]{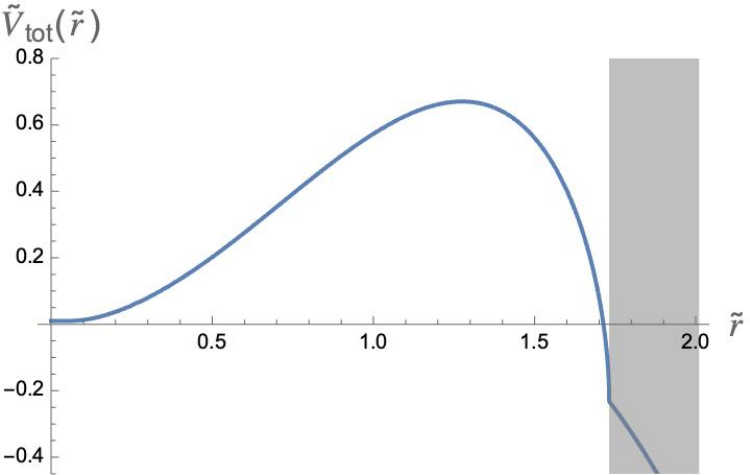}
\subcaption{Metastable potential for $\widetilde{b}_{D3}=0.01$ and $\widetilde{G}_{4}=0.5$. The shell radius is $r_{\rm min}\approx 0.0037$.
}
\end{minipage}
\hfill
\begin{minipage}[t]{0.5\hsize}
\centering
\includegraphics[width=0.9\linewidth]{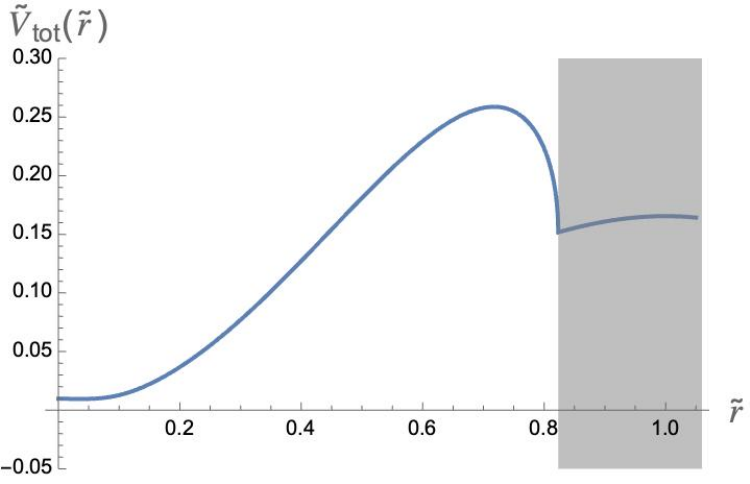}
\subcaption{Stabilized potential for $\widetilde{b}_{D3}=0.01$ and $\widetilde{G}_{4}=2.2$. The shell radius is $r_{\rm min}\approx 0.72$.
}
\end{minipage}
\caption{Brane shell potentials for each parametrization. Shaded regions represent the outside of the horizon.
}
\label{braneShell-someParameter}
\end{figure}

Let us assume we have a potential minimum $V(r_{\rm min})$. The probe level potential \eqref{probPot} is an unbounded potential like a cubic anharmonic oscillator. However, as mentioned in the previous section, we need to consider the horizon to study the stability of the potential that gravitational correction is included. Here, the concrete shape of the corrected potential is determined by $G_{4}$ and $b_{D3}$, and we can plot it as Figure \ref{braneShell-someParameter} under some parametrization. The dimensionless potential of the brane shell is given by 
\begin{align}
    &V_{\rm tot}=2\pi T_{DW}c^{2}\widetilde{V}_{\rm tot}\ , \label{dimlessVtot1} \\
    &\widetilde{V}_{\rm tot}=\sqrt{\widetilde{r}^{4}+\widetilde{b}^{2}_{D3}}\cdot \frac{1}{2}\left(1+\sqrt{1-\frac{2\widetilde{G}_{4}\widetilde{m}(\widetilde{r})}{\widetilde{r}}} \right)-\widetilde{b}_{D3}\widetilde{r}{}_{2}F_{1}\left(-\frac{1}{2},\frac{1}{4},\frac{5}{4};-\frac{\widetilde{r}^{4}}{\widetilde{b}_{D3}^{2}} \right) ,\label{dimlessVtot2}
\end{align}
where $c=T_{DW}/2T_{D5}r_{NS}$. In this point, we introduced the dimensionless mass function and dimensionless Newton constant as
\begin{align}
&m(r) = 2\pi T_{DW}c^{2}\widetilde{m}(\widetilde{r}),\quad \widetilde{m}(\widetilde{r}) = \widetilde{b}_{D3}\widetilde{r}{}_{2}F_{1}\left(-\frac{1}{2},\frac{1}{4},\frac{5}{4};-\frac{\widetilde{r}^{4}}{\widetilde{b}_{D3}^{2}} \right), \label{dimlessMassfunc} \\
&G_{4} = c^{\prime}\widetilde{G}_{4},\quad c^{\prime} = \frac{T_{D5}r_{NS}}{\pi T_{DW}^{2}}. \label{dimlessG}
\end{align}
In the potential shown in Figure \ref{braneShell-someParameter}-(a), the potential value at the horizon is lower than the minimum. In this situation, the brane shell is metastable and then can decay via quantum tunneling in a finite lifetime. In contrast, in Figure \ref{braneShell-someParameter}-(b), where we fix the dimensionless Newton constant to a slightly large value,  the value of the potential at the horizon is larger than the minimum conversely. This difference can be interpreted as the shell being stabilized due to the gravitational effect.

Summarizing the conditions, whether the system is stable or metastable is distinguished by whether or not the potential value at the horizon is larger than $V(r_{\rm min})$. That is, 
\begin{align}
    & V(r_{\rm min}) \geq V(r_{\rm ch})\quad \text{metastable}, \\
    & V(r_{\rm min}) < V(r_{\rm ch})\quad \text{stable}.
\end{align}
It is difficult to write down $V(r_{\rm min})$ explicitly because it has a complicated expression that includes a hypergeometric function. On the other hand, the value of the potential at the horizon can be rewritten in a somewhat simpler expression. Since $f_{+}(r_{\rm ch})=0$ at the horizon, the potential value is as follows from \eqref{totalEnergy_rev}
\begin{align}
    V(r_{\rm ch}) = \pi T_{\rm DW}\sqrt{r^{4}_{\rm ch}+b_{D3}^{2}} -4\pi T_{D5}r_{NS} \int^{r_{\rm ch}}_{0}d\xi \sqrt{\xi^{4}+b^{2}_{D3}}. \label{pot_ch}
\end{align}
We can rewrite the second term using $f_{+}(r_{\rm ch})=0$, namely
\begin{align}
    &r_{\rm ch} = 2G_{4}\cdot 4\pi T_{D5}r_{NS} \int^{r_{\rm ch}}_{0}d\xi \sqrt{\xi^{4}+b^{2}_{D3}} \nonumber \\
    &\hspace{0.5cm} \longrightarrow 4\pi T_{D5}r_{NS} \int^{r_{\rm ch}}_{0}d\xi \sqrt{\xi^{4}+b^{2}_{D3}} = \frac{r_{\rm ch}}{2G_{4}}.
\end{align}
Then, the potential value becomes
\begin{align}
V(r_{\rm ch}) = \pi T_{\rm DW}\sqrt{r^{4}_{\rm ch}+b_{D3}^{2}} - \frac{r_{\rm ch}}{2G_{4}},
\end{align}
and the simplest expression of the stability condition is
\begin{align}
    & V(r_{\rm min})  \geq \pi T_{\rm DW}\sqrt{r^{4}_{\rm ch}+b_{D3}^{2}} - \frac{r_{\rm ch}}{2G_{4}}\quad \text{metastable}, \\
    & V(r_{\rm min})  < \pi T_{\rm DW}\sqrt{r^{4}_{\rm ch}+b_{D3}^{2}} - \frac{r_{\rm ch}}{2G_{4}}\quad \text{stable}.
\end{align}

\begin{figure}[tb]
\begin{minipage}[t]{0.5\hsize}
\centering
\includegraphics[width=0.9\linewidth]{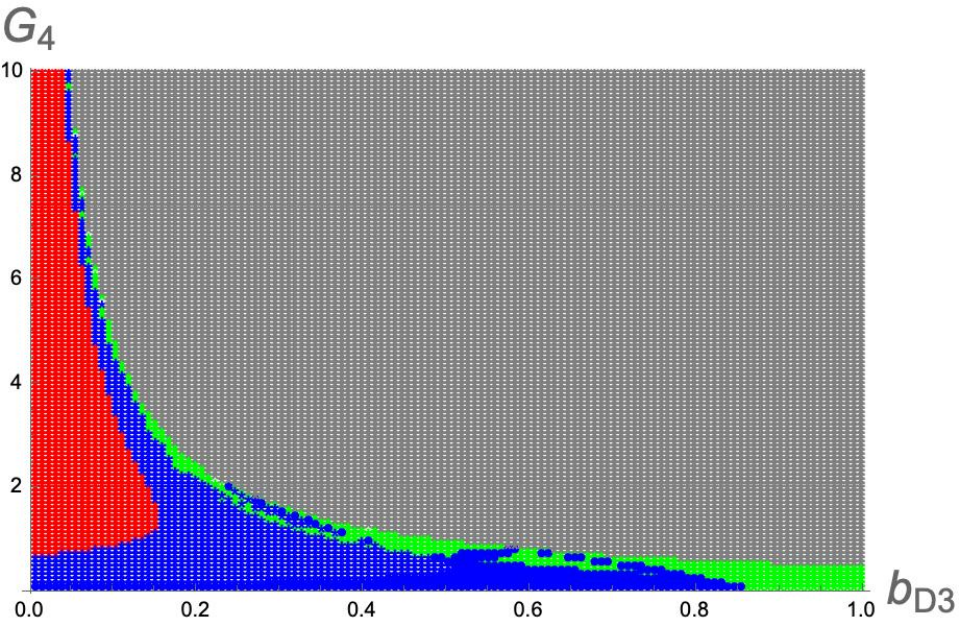}
\subcaption{}
\end{minipage}
\hfill
\begin{minipage}[t]{0.5\hsize}
\centering
\includegraphics[width=0.9\linewidth]{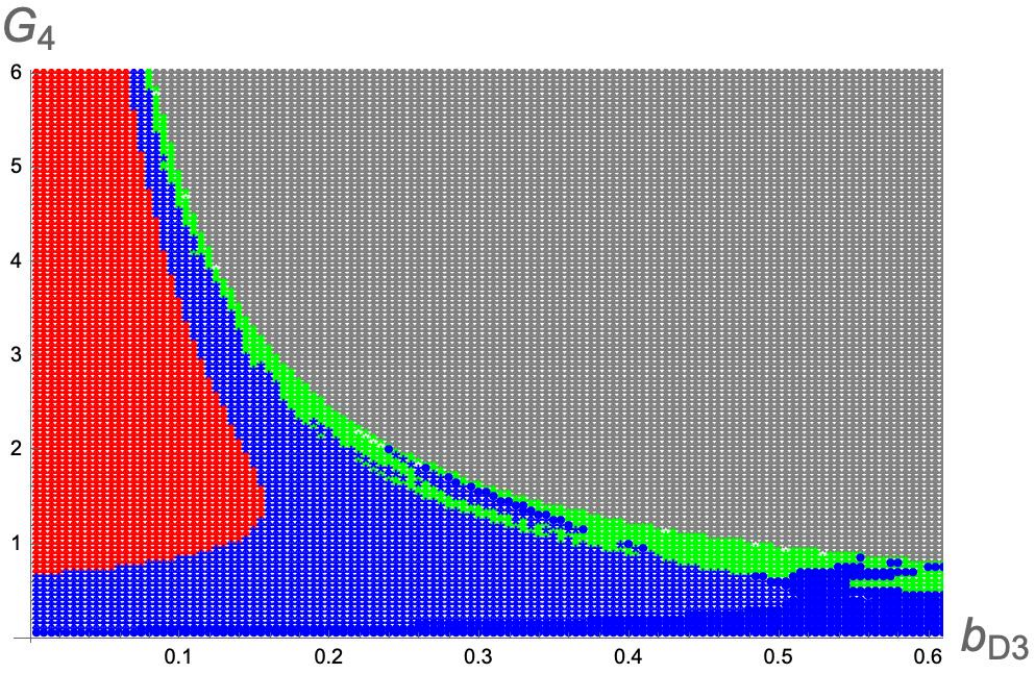}
\subcaption{}
\end{minipage}
\caption{Blue area represents a metastable region, red area represents a region where the shell is stabilized, green area represents a region where the potential is monotonically decreasing and there is no minimum, and gray area represents a region where $f_{+}(r)<0$ for all $r$ and no physical spacetime exists. Note that axes is labeled by dimensionless variables. The blue areas scattered throughout the green area are due to numerical errors, and we cannot actually constitute a stable shell at those point.
}
\label{shell-phaseDiagram}
\end{figure}

Based on this condition, we can confirm the brane shell's state for different magnetic field values and the Newton constant numerically, and the state plot is as in Figure \ref{shell-phaseDiagram}-(a). As it is naively expected that the stronger the gravitational correction is, the more stable the shell is, stable shells would be realized with a somewhat large value of the Newton constant. However, if we fix the magnetic field to any value larger than about 0.15, no stabilized state exists, no matter how large the Newton constant is. This means that we cannot stabilize brane shells for a sufficiently strong magnetic field even if considering the gravitational correction. Moreover, any number of D3-branes could be wrapped at the probe level, but in the present discussion with gravitational corrections, if the magnetic field is too strong, $f_{+}$ becomes always negative, and a physical spacetime would no longer exist. As we mentioned in the section \ref{Review of bound state between domain wall D5 and monopole}, the value of the magnetic field is proportional to the number of D3-branes. Thus, it suggests the existence of the maximum number of D3-branes wrapped into the internal space, which can not be confirmed in \cite{Danielsson:2022odq}.

\subsection{Over-extremality}
\label{Over-extremal property}

In addition to discussing the mechanism by which the brane shell is stabilized through gravitational corrections, we are also interested in whether the stabilized shell can become over-extremal from the perspective of the swampland conjectures. The over-extremality of compact objects is defined by the weak gravity condition \eqref{WGcond-original} \cite{Arkani-Hamed:2006emk}. This condition roughly states that gravity is the weakest force and becomes trivial when taking a limit as $M_{\rm pl}\rightarrow\infty$. Deforming the condition, we obtain
\begin{align}
    \frac{G_{4}m^{2}}{(gQ)^{2}}\leq 1\ . \label{WGcond3}
\end{align}
Note that we absorb the $U(1)$ coupling to the definition of the charge in our model as $gQ=q$. In addition, we can rewrite this $q$ by the magnetic field coming from D3-branes. Then, the inequality \eqref{WGcond3} becomes
\begin{align}
    \frac{G_{4}m^{2}}{(gQ)^{2}}=\frac{G_{4}m^{2}}{q^{2}}=\frac{G_{4}m^{2}}{\gamma^{-1}b^{2}_{D3}}\leq 1\ , \label{WGcond-rev}
\end{align}
where $m$, which is the mass of the brane shell, is given as the minimum of the brane shell potential \eqref{totalEnergy_rev} \cite{Danielsson:2022odq}
\begin{align}
    \partial_{r}V(r_{\rm min})=0\ ,\quad V(r_{\rm min})=m\ .
\end{align}
For the numerical evaluation, we also show the dimensionless version of \eqref{WGcond-rev} 
\begin{align}
	\frac{4\pi\widetilde{G}_{4}\widetilde{m}}{\widetilde{b}_{D3}^{2}} \leq 1, \label{dimensionlessWGcond}
\end{align}
where we use the dimensionless parameters introduced in \eqref{dimlessVtot1}-\eqref{dimlessG}.

Figure \ref{shell-phaseDiagram}-(b) is an enlarged view of the region where $G_{4}$ and $b_{D3}$ are small. The light blue region plotted with stars ({\color{blue}$\star$}) represents non-extremal metastable states. In contrast, the deep blue region plotted with circles ({\color{blue}$\bullet$}) represents over-extremal states. The red region is entirely indicated by star symbols, showing that there are no over-extremal stabilized shells. This fact is consistent with the difficulty of satisfying the weak gravity condition as the dimensionless Newton constant becomes larger.

\begin{figure}[tb]
    \centering
    \includegraphics[width=0.6\linewidth]{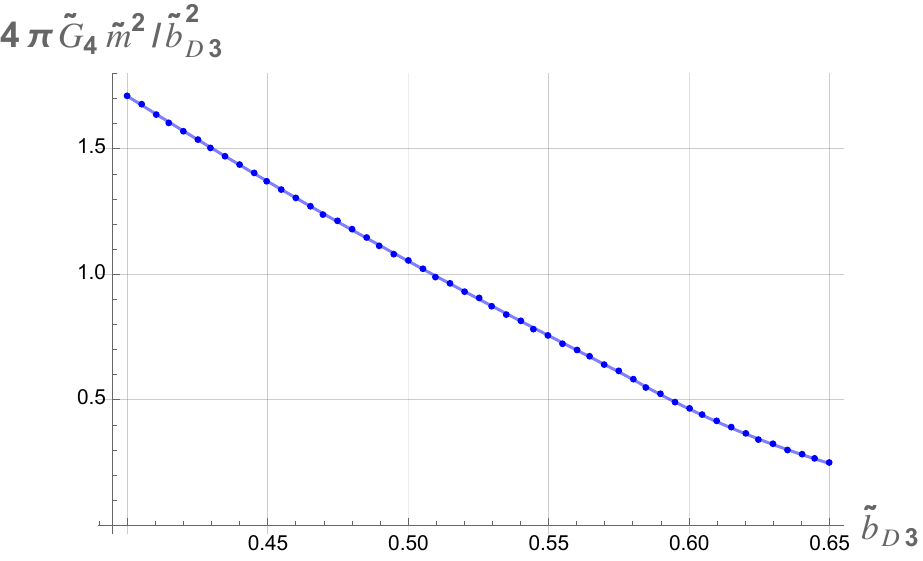}
    \caption{Mass to charge ratio at $\widetilde{G}_{4}=0.5$. It can be seen that the brane shell becomes over-extremal when the magnetic field is larger than about $\widetilde{b}_{D3}=0.51$. 
}
    \label{mqratio}
\end{figure}

Let us consider a situation where the magnetic field is increased with a fixed dimensionless Newton constant. Here, we fix the dimensionless Newton constant to an arbitral value less than 1. As seen from Figure \ref{mqratio}, whereas the brane shell is non-extremal in somewhat small magnetic field regions, it becomes over-extremal at some point as the magnetic field is increased. This behavior is easily expected from the weak gravity condition \eqref{WGcond-rev} (or \eqref{dimensionlessWGcond}). However, if the magnetic field is too strong, the potential minimum disappears, and (meta)stable shells cannot exist. If the magnetic field is made even stronger, there is no physical region at all. Although there was no constraint on the magnitude of the charge of the shell, i.e., the number of D3-branes in \cite{Danielsson:2022odq}, in our setup, it is impossible to make the magnetic field arbitrarily large to achieve an over-extremal state under the gravitational correction, and there exists a certain upper bound. For each Newton constant, the upper bound on the magnetic field for at least the metastable shell to exist is characterized as the boundary between the green and blue regions.

\section{Discussion and summary}
\label{Discussion and summary}
In this paper, we geometrically realized a metastable state with D5-branes and anti D5-branes and considered the situation that a two-sphere of domain wall D5-brane, i.e., brane shell emerges in 4D spacetime. Here, we wrapped D3-branes in the internal space so that the shell magnetically charged via D3-brane dissolving exists, as seen from an observer in 4D spacetime. This shell has a finite radius even before tunneling due to the nonlinear nature of the DBI action. We added a gravitational correction by the junction condition and discussed the stability of the brane shell and the over-extremality in the sense of the weak gravity conjecture. Although the gravitational correction does not change the shape of the potential itself, a ``cosmological'' horizon appears in the spacetime outside the shell. The position of the horizon is determined by the values of the background magnetic field $b_{D3}$ and 4D Newton constant $G_{4}$. We have numerically shown that the shell can be stabilized at a finite radius due to the gravitational correction, especially in situations where the Newtonian constant is large enough. On the other hand, when we fixed the Newton constant at an arbitrary value and increased the magnetic field, it was confirmed that the brane shell could be an over-extremal state that satisfies the so-called weak gravity condition in the parameter region where the magnetic field is large.

Our model could not identify any state of the brane shell that is fully stabilized and over-extremal under the gravitational correction. It implies that there is some kind of dilemma between taking a large Newton constant to stabilize the shell and taking a large magnetic field to satisfy the weak gravity condition. We can read this tendency from the expression of the weak gravity condition \eqref{WGcond-original} (or \eqref{WGcond3}). In particular, in the case of our model, if we make both the Newton constant and the magnetic field too large, the potential minima and the physical space-time itself will disappear, so we cannot make both parameters very large and achieve both stability and over-extremality. Whether it is possible to construct an over-extremal and completely stable state in a stringy model is an intriguing problem we would like to address in future work.

\section*{Acknowledgments}
We want to express profound gratitude to Yutaka Ookouchi for fruitful discussions and help. This work is supported by the Kyushu University Leading Human Resources Development Fellowship Program.

\end{document}